\documentclass[referee]{aa}
\usepackage{graphicx,natbib}

\title{On the composition of ices incorporated in Ceres}

\author{Olivier Mousis  \and Yann Alibert}

\institute{
Physikalisches Institut, University of Bern, Sidlerstrasse 5, CH-3012 Bern, Switzerland \\
}

\begin{document}

\date{Accepted 2004 December 22. Received 2004 November 03; in original form 2004 September 02}

\label{firstpage}

\abstract
{We use the clathrate hydrate trapping theory (Lunine~\&~Stevenson~1985) and the gas drag formalism of Weidenschilling~(1977) to calculate the composition of ices incorporated in the interior of Ceres. In the spirit of the work of Cyr et al. (1998), and using a time dependent solar nebula model, we show that icy solids can drift from beyond 5 AU to the present location of the asteroid and be preserved from vaporization. We argue that volatiles were trapped in the outer solar nebula under the forms of clathrate hydrates, hydrates and pure condensates prior to having been incorporated in icy solids and subsequently in Ceres. Under the assumption that most of volatiles were not vaporized during the accretion phase and the thermal evolution of Ceres, we determine the per mass abundances with respect to H$_2$O of CO$_2$, CO, CH$_4$, N$_2$, NH$_3$, Ar, Xe, and Kr in the interior of the asteroid. The $Dawn$ space mission, scheduled to explore Ceres in August 2014 (Russel et al. 2004), may have the capacity to test some predictions. We also show that an $in~situ$ measurement of the D/H ratio in H$_2$O in Ceres could  constrain the distance range in the solar nebula where its icy planetesimals were produced.


}

\titlerunning{Composition of ices in Ceres}
\authorrunning{Mousis O. \& Alibert Y.}

\maketitle

\section{Introduction}

The C-class asteroid Ceres was discovered in 1801 by the Italian astronomer Giuseppe Piazzi. Its mass, estimated to be about $(4.70 \pm 0.04) \times 10^{-10}$ $M_{\odot}$ (Michalak 2000), represents 30 to 40 \% of that of the main belt. Ceres possesses several intriguing physical characterictics. Among them, a density estimated to be about $2.03  \pm 0.05$ g.cm$^{-3}$ (Michalak 2000, Parker et al. 2002) close to that of some existing carbonaceous chondrite meteorites, as well as of Ganymede and Callisto, combined with a low porosity, suggest that a large amount of volatiles was trapped in the interior of this asteroid during its accretion (Russel et al. 2004). Note that this low density does not necessarily imply the presence of ices in Ceres. On the other hand, several arguments tend to support this assumption. Thus, McCord~\&~Sotin (2003) estimated that the mass fraction of incorporated ices is about 20 \% of the total mass of Ceres. This hypothesis is in agreement with the recent HST observations conducted by Parker et al. (2004) who showed, from measurements of Ceres oblateness, that its internal structure is compatible with a body containing a central rocky core surrounded by an icy shell. Moreover, the presence of volatiles in Ceres is consistent with the detection of OH escaping from a north polar icy cap that might be replenished during winter by subsurface percolation, but which dissipates in summer (A'Hearn~\&~Feldman~1992). In addition, due to its large size ($\sim$ 950 km diameter), Ceres is expected  to have experienced some processes (differentiation, preservation of a deep liquid layer) normally associated with the evolution of major icy satellites (McCord~\&~Sotin 2003). In this work, we then favor the hypothesis of the presence of important quantities of ices in the interior of Ceres to explain all the afore-mentioned evidences. \\

In order to estimate the distribution of water in the solar nebula, Cyr~et~al~(1998) utilized the stationary solar nebula models of Cassen (1994) and showed that icy particles originating from beyond 5 AU may litter the 3 to 5 AU zone by drifting inward. This result may be confirmed by the high porosity observed in some asteroids that would result from the escape of the accreted volatiles during aqueous alteration (Wilson et al. 1998). Hence, asteroids located between 3 and 5 AU and initially including a high proportion of ices would have been accreted from a mix of rocky planetesimals formed locally and icy planetesimals originating from distances farther than 5 AU in the solar nebula. However, the efficiency of the mechanism proposed by Cyr~et~al.~(1998) remains uncertain to explain the potential incorporation of volatiles in Ceres, assuming it formed at its current orbit (2.7 AU).\\

In this work, we focus on the possibility of determining the nature and the composition of ices that may be included in the interior of Ceres by reconsidering the scenario of Cyr et al. (1998) and by using the clathrate hydrate trapping theory described by Lunine \& Stevenson (1985). Using a time dependent model of the solar nebula, we first show that icy planetesimals may drift from heliocentric distances larger than 5 AU to the present location of Ceres without encountering temperature and pressure conditions high enough to vaporize the ices they contain. We then constrain the composition of ices incorporated in these planetesimals and thus that of ices included in Ceres. Such a study may improve our knowledge of the internal structure of Ceres since interior models strongly depend on the assumed composition. The $Dawn$ space mission, which is scheduled to study the geophysics and the geochemistry of Ceres in detail during its approach in August 2014 (Russel et al. 2004), may have the capacity to test some predictions. Furthermore, in order to constrain the origin of icy solids that were incorporated in Ceres, we consider the work of Mousis (2004) who calculated the range of plausible values of the D/H ratio in H$_2$O ice as a function of the heliocentric distance in the solar nebula. We show that future measurements of the D/H ratio in H$_2$O in Ceres should allow the distance range in the solar nebula where the icy planetesimals were formed to be determined.\\

In Sect. 2, we calculate the radial drift of icy solids, as a result of gas drag, in a time dependent solar nebula model. In Sect. 3, we examine the thermodynamical conditions needed for the trapping of major volatiles in the outer solar nebula. We then calculate the values of the per mass abundances of these species with respect to H$_2$O in icy planetesimals formed in the outer solar nebula and subsequently accreted in Ceres. In Sect. 4, we study the possibility of constraining the origin of these planetesimals in the solar nebula as a function of their D/H ratio in H$_2$O. Sect. 5 is devoted to summary and discussion.

\section{Orbital decay of icy particles}

 \subsection{Disk model}
 Once formed, planetesimals suffer orbital decay due to gas drag (Weidenschilling 1977). As a result, icy particles formed in the outer solar system may have migrated to inner regions, then bringing some material trapped at higher distances from the Sun. This has been shown in particular by Cyr et al. (1998) who calculated the evolution of icy particles due to gas drag, sedimentation and sublimation. They showed that some particles originating from $\sim 5$ AU may migrate to heliocentric distances of the order of 3 AU due to gas drag, before being sublimated due to higher temperature and pressure domains in the inner solar system. Hence, these icy particles may be further incorporated in bodies formed at these locations. However, the innermost heliocentric radius of these particles (3 AU) calculated by Cyr et al. (1998) is too large to allow their incorporation in Ceres, which is located at $\sim 2.7$ AU. On the other hand, these calculations are based on static solar nebula models by Cassen (1994). In an evolving nebula, due to viscosity, the surface density and the temperature decrease as a function of time. The particles may actually survive to heliocentric distances shorter than first estimated by Cyr~et~al.~(1998).\\

 In order to quantify this effect, we have performed calculations similar to the ones of Cyr et al. (1998). We use for the gas drag the Weidenschilling (1977) formalism, and an evolving solar nebula model taken from Alibert et al (2004a). In this model, cylindrical symmetry is assumed and we use the cylindrical coordinates $(r,z,\theta)$, where $r$ is the distance to the Sun, $z$ the height and $\theta$ the azimutal coordinate. The disk is assumed to be thin.
 The surface density $\Sigma$ in the disk is calculated by solving the diffusion equation:

 \begin{equation}
 {d \Sigma \over d t} = {3 \over r} {\partial \over \partial r } \left[
 r^{1/2} {\partial \over \partial r}
 ( \tilde{\nu} \Sigma r^{1/2})  \right]  + \dot{\Sigma}_w(r)
 ,
 \end{equation}

 where $\tilde{\nu}$ is the mean viscosity (integrated along the $z$ axis), and $\dot{\Sigma}_w(r)$ is the photoevaporation term, taken as in Veras \&
 Armitage (2004). The viscosity is calculated in the framework of the $\alpha$-formalism (Shakura \& Sunyaev 1973) using the following method (see Alibert~et~al.~2004a for details). For each distance to the Sun $r$, we calculate the vertical structure of the disk, by solving the equations of hydrostatic equilibrium, radiative diffusion and energy balance between viscous dissipation and radiative losses. This gives us $\tilde{\nu}$, as well as $P_{\rm
 disk}$ and $T_{\rm disk}$ the disk midplane pressure and temperature, as a function of the surface density $\Sigma$.\\

 We start the evolution of the disk with an initial profil given by $\Sigma \propto r^{-3/2}$, the constant been chosen to have a disk mass of $\sim 0.04$ $M_{\odot}$ (between 0.5 and 50 AU). The total rate of mass loss by photoevaporation is $10^{-8} M_{\odot}$/yr, and the $\alpha$ parameter is taken equal to $2 \times 10^{-3}$. This model is similar to the one that can lead to the formation of a Jupiter like planet (see Alibert et al. 2004b).

\subsection{Radial drift}
 We start our radial drift calculations at different times $t_i$ (ranging from 0 to $\sim 3$ Myr, the lifetime of the disk)  with particles initially located at $r_i$ = 5 to 15 AU. Since we are interested by the content in volatile species of these icy particles, we consider only particles that start their drift when the temperature $T_{\rm disk}$ has fallen under $38$ K (the lowest clathration temperature of the species we consider in this work, see Fig. 2).\\

 Figure 1 represents the initial parameters $r_i$ and $t_i$ for particles that drift to heliocentric distances lower than 2.7 AU, and that have encountered
 a disk temperature below $38$ K between $r_i$ and 2.7 AU. This figure shows that particles formed between 5 and 15 AU can survive to heliocentric distances low enough to be incorporated in Ceres during its accretion phase. Moreover, since the temperature of gas around the particles remains low, their content in volatile species is not modified between their formation and their incorporation in the asteroid during its accretion phase. In lower mass asteroids, the volatiles may subsequently be lost, thus explaining their high porosity (Wilson et al. 1999).\\

 Note that, in these simulations, contrary to the one of Cyr et al. (1998), the radius of particles is assumed to be constant: sublimation and particle growth are not considered. However, since the surrounding temperature remains always below $38$ K, sublimation can be neglected. Concerning particles growth, this effect could modify the drift velocity and extent, but, as shown by Fig. 1, our conclusions do not depend strongly on the particle size, provided it ranges between 0.1~m and 10 m, the sizes we have considered.\\

 Finally, the calculation of the disk structure provides us the cooling curve that is used in the following Section to determine the composition of ices utimately incorporated in Ceres.

\begin{figure}
\includegraphics[width=70mm]{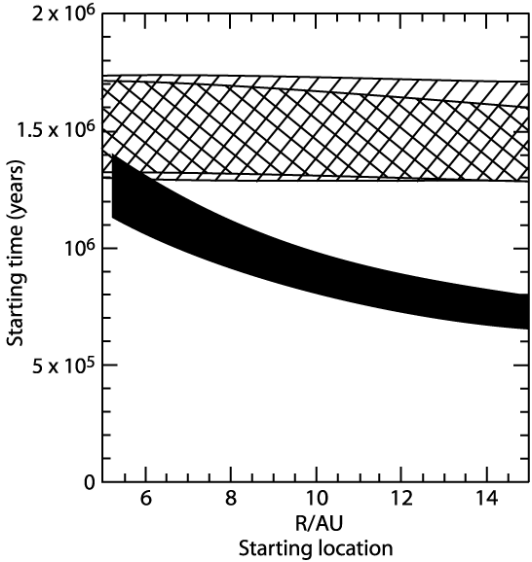}
\caption{Initial parameters ($t_i$ and $r_i$, see text) of icy particles that reach the 2.7 AU line without any loss of trapped volatiles. The $///$ and $\backslash \backslash \backslash$ dashed areas correspond to particles with sizes of 0.1 and 1 m, respectively. The dark area corresponds to particles with size of 10 m.}
\label{Ceres_fig1}
\end{figure}

\section{Composition of ices incorporated in Ceres}
   \subsection{Formation of ices in the outer solar nebula}
As we demonstrated in Section 2, during its accretion phase, Ceres may have incorporated icy solids that were formed beyond 5 AU in the solar nebula. Such a scenario leads us to examine the thermodynamical conditions required for the trapping of volatiles in the outer solar nebula. We argue that volatiles were trapped under the form of hydrates, clathrate hydrates or pure condensates prior to have been integrated in icy planetesimals. Figure 2 illustrates the trapping process considered with stability curves for clathrate hydrates of major species calculated from the thermodynamical data of Lunine \& Stevenson (1985) and with solar nebula cooling curves located at 5 and 15 AU taken from the model described in Section 2. The stability curve of CO$_2$ pure condensate is derived from the existing experimental data (Lide 1999). From Fig. 2, it can be seen that CO$_2$ crystallizes as a pure condensate prior to be trapped by water to form a clathrate hydrate during the cooling of the solar nebula. Hence, we assume in this work that solid CO$_2$ is the only existing condensed form of CO$_2$ in the solar nebula. Species remain in the vapor phase as long as they stay in the domains located above the curves of stability. The intersections of the cooling curves with the stability curves determine the thermodynamical conditions (temperature $T$, pressure $P$ and surface density $\Sigma$) at which the different condensates were formed at 5 and 15 AU. Once formed, the different ices agglomerated and were incorporated in icy planetesimals. As we discuss in previous Section, these icy solids drifted inward and a part of them was accreted by asteroids during their formation.

\subsection{Relative abundances of volatile compounds in Ceres}
In this work, we assume that abundances of all elements are solar (Anders \& Grevesse 1989) and that O, C, and N exist only under the form of H$_2$O, CO$_2$, CO, CH$_4$, N$_2$, and NH$_3$ in the solar nebula gas-phase. Moreover, we follow the study of Mousis~et~al.~(2002a) who showed that CO/CH$_4$ and N$_2$/NH$_3$ ratios in the solar nebula gas-phase remain quasi identical to the interstellar values by taking into account turbulent diffusion and the kinetics of chemical conversions. According to their work, the CO/CH$_4$ adopted ratio in the solar nebula must be of the order of the measured ISM value. Hence, we adopt CO/CH$_4$~=~10 in the solar nebula gas-phase in all our calculations, a value derived from the ISM measurements of Allamandola~et~al.~(1999). Moreover, since large amounts of CO$_2$ have been detected in ISM ices (Gibb et al. 2004), we also assume the presence of this species in the solar nebula gas-phase with an abundance close to the ISM value. We then consider CO$_2$/CO = 3 in the solar nebula gas-phase, a value consistent with the ISM measured one reported by Gibb et al. (2004). On the other hand, the question of the N$_2$/NH$_3$ ratio is still open since, until now, no constraining measurements have been made in the interstellar medium. Here, we consider the values of 0.1, 1, and 10 for this ratio. We also assume, as in Gautier et al. (2001a,b) and Alibert et al (2004b), that the amount of water available in the solar nebula is at least high enough to allow the trapping of the considered volatile species, except CO$_2$, in the form of hydrates or clathrate hydrates. This water enhancement may result from the inward drift of icy grains, and from local accumulation of water vapor at radii interior to the evaporation/condensation front, as described by Cuzzi \& Zahnle (2004). The corresponding minimum molar mixing ratio of water relative to H$_2$ in the solar nebula gas-phase is given by

\begin{equation}
{x_{H_2O} = \sum_{\it{i}} \gamma_i~x_i~\frac{\Sigma(R; T_i, P_i)}{\Sigma(R; T_{H_2O}, P_{H_2O})}},
\end{equation}

where $x_i$ is the molar mixing ratio of the volatile $i$ with respect to H$_2$ in the solar nebula gas-phase, $\gamma_i$ is the required number of water molecules to form the corresponding hydrate or clathrate hydrate (5.75 for a type I hydrate, 5.66 for a type II hydrate, 1 for the NH$_3$-H$_2$O hydrate and 0 for CO$_2$ pure condensate), $\Sigma(R; T_i, P_i)$ and $\Sigma(R; T_{H_2O}, P_{H_2O})$ are the surface density of the nebula at the distance $R$ from the Sun at the epoch of hydratation or clathration of the species $i$ and at the epoch of condensation of water, respectively. The resulting minimum abundance of water required at 5 AU is 2.13, 2.03, and 1.88 times the water abundance in the solar nebula (H$_2$O/H$_2$ = $4.86 \times 10^{-4}$ with CO$_2$/CO/CH$_4$~=~30/10/1) for values of N$_2$/NH$_3$ = 10, 1, and 0.1 in the solar nebula gas-phase, respectively.\\

For a given abundance of water in the solar nebula, the mass ratio of the volatile $i$ to water in icy planetesimals formed at a distance $R$ from the Sun is determined by the relation derived from Mousis~\&~Gautier~(2004), which is

\begin{equation}
{Y_i = \frac{X_i}{X_{H_2O}} \frac{\Sigma(R, t_i)}{\Sigma(R, t_{H_2O})}},
\end{equation}

where $X_i$ and $X_{H_2O}$ are the mass mixing ratios of the volatile $i$ and of H$_2$O with respect to H$_2$ in the solar nebula, respectively. Table I displays the values of the mass ratios of the different volatiles to H$_2$O  in planetesimals formed at 5 AU for N$_2$/NH$_3$~=~10, 1, and 0.1 in vapor phase in the solar nebula and for the calculated minimum abundances of water. Similar calculations have been done for planetesimals formed at 10, 15 and 20 AU and reveal that mass ratios of the volatiles to H$_2$O remain almost constant whatever the distance to the Sun (see Alibert et al. 2004b for details).\\

Assuming that 1) icy planetesimals accreted by Ceres were produced in this distance range and 2) most of the trapped volatiles were not lost during the formation and the thermal history of the asteroid, the per mass abundances of these species in the interior of the current Ceres can be deduced from Table I. Hence, it can be seen that in Ceres, the CO$_2$/H$_2$O, CO/H$_2$O and CH$_4$/H$_2$O ratios should be roughly equal to 90 wt\%, 14 wt\% and 9 wt\%, respectively. Moreover, depending on the assumed ISM N$_2$/NH$_3$ value, the N$_2$/H$_2$O and NH$_3$/H$_2$O ratios should be between 1.5 wt\% and 8 wt\%, and 0.7 wt\% and 14.3 wt\%, respectively. In addition, H$_2$S should be present in Ceres with a H$_2$S/H$_2$O ratio of the order of 4.5 to 5 wt\%. Low amounts of Ar, Kr, and Xe should also exist in the interior of the asteroid.

\begin{figure}[h]
\centering
\includegraphics[width=6cm,clip=,angle=-90]{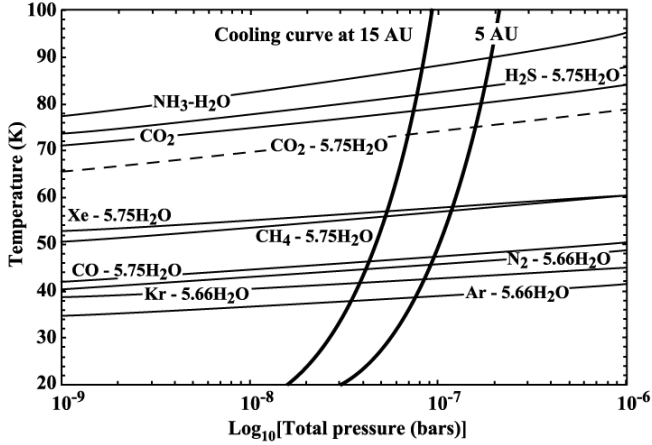}
\caption{Stability curves $T = f(P)$ of species trapped as hydrates or clathrate hydrates considered in this work and evolutionary tracks of the solar nebula in $P - T$ space at the heliocentric distances of 5 and 15 AU. Abundances of various elements are solar with CO$_2$/CO/CH$_4$~=~30/10/1 and N$_2$/NH$_3$~=~1 in vapor phase. The condensation curve of CO$_2$ pure condensate (solid line) is plotted together with that of the corresponding clathrate hydrate (dashed line). Evolutionary tracks at 5 and 15 AU are derived from the solar nebula model described in Section 2.}
\label{FigVibStab}
\end{figure}

	 \begin{table}
            \caption[]{Calculations of the ratios of trapped masses of volatiles to the mass of H$_2$O ice in planetesimals formed at 5 AU in the solar nebula. Gas-phase abundance of H$_2$O is determined with Eq. 1 and gas-phase abundances of volatiles are assumed to be solar (Anders~\&~Grevesse~1989) with CO$_2$/CO/CH$_4$~=~30/10/1 and N$_2$/NH$_3$~=~0.1, 1 or 10 in the solar nebula gas-phase.}
      \begin{center}
         {\setlength{\tabcolsep}{0.1cm}
         \begin{tabular}{lccc}
	    \hline
            \hline
            \noalign{\smallskip}
	      & N$_2$/NH$_3$~=~0.1  & N$_2$/NH$_3$~=~1  & N$_2$/NH$_3$~=~10  \\
             \noalign{\smallskip}
             \hline
             \noalign{\smallskip}
	      CO$_2$/H$_2$O   &  $9.65 \times 10^{-1}$  &  $8.93 \times 10^{-1}$  &  $8.54 \times 10^{-1}$   \\
	      CO/H$_2$O  	 &  $1.51 \times 10^{-1}$  &  $1.39 \times 10^{-1}$  &  $1.33 \times 10^{-1}$   \\
	      CH$_4$ /H$_2$O  &  $9.67 \times 10^{-3}$  &  $8.94 \times 10^{-3}$  &  $8.55 \times 10^{-3}$   \\
	      N$_2$/H$_2$O     &  $1.53 \times 10^{-2}$  &  $5.82 \times 10^{-2}$  &  $8.01 \times 10^{-2}$    \\
	      NH$_3$/H$_2$O   &  $1.43 \times 10^{-1}$  &  $5.23 \times 10^{-2}$  &  $6.91 \times 10^{-3}$   \\
	      H$_2$S/H$_2$O   &  $5.11 \times 10^{-2}$  &  $4.72 \times 10^{-2}$  &  $4.52 \times 10^{-2}$   \\
	      Ar/H$_2$O    	 &  $5.91 \times 10^{-3}$  &  $5.47 \times 10^{-3}$  &  $5.23 \times 10^{-3}$   \\
	      Kr/H$_2$O    	 &  $3.06 \times 10^{-6}$  &  $2.83 \times 10^{-6}$  &  $2.71 \times 10^{-6}$    \\
	      Xe/H$_2$O    	 &  $3.90 \times 10^{-7}$  &  $3.61 \times 10^{-7}$  &  $3.45 \times 10^{-7}$    \\
           \noalign{\smallskip}
            \hline
         \end{tabular}}
       \end{center}
      \end{table}

   \section{Constraining the origin of Ceres by measuring the D/H ratio in H$_2$O}
Understanding the formation conditions of Ceres requires observational tests to be provided that can constrain the origin of the low density and volatile rich planetesimals that led to its accretion. The study of the evolution of the D/H ratio in H$_2$O in the solar nebula can provide such a constraint since the rate of the isotopic exchange between HDO and H$_2$ in vapor phase strongly depends on local conditions of temperature and pressure which, in turn, depend on the heliocentric distance and the temporal evolution of the solar nebula. Taking into account the fact that the isotopic exchange is inhibited when H$_2$O condenses, this requires that the value of the D/H ratio in microscopic ices is the one at the time and at the location of the condensation of vapor. When these icy grains reached millimeter size, they started to decouple from gas, continued to grow and formed planetesimals. Whatever the subsequent evolution of these solids, their D/H ratio is that of microscopic icy grains.\\

In order to constrain the range of plausible values of the D/H ratio in H$_2$O ice included in Ceres, we refer to the works of Drouart et al. (1999), Mousis et al. (2000) and Mousis (2004) who described the evolution of the D/H ratio in H$_2$O in the solar nebula as a function of time and the distance from the Sun. Drouart et al. (1999) employed the time dependent turbulent model developed for the solar nebula by Dubrulle (1993) in order to interpret the measurements of the D/H ratio in LL3 meteorites, and in comets P/Halley, C/1996 B2 (Hyakutake), and C/1995 O1 (Hale-Bopp). This model depends on three physical parameters: the initial mass of the nebula $M_{D0}$, its initial radius $R_{D0}$, and the coefficient of turbulent viscosity $\alpha$, derived from Shakura \& Sunyaev (1973). Drouart et al. (1999) calculated the deuterium enrichment factor $f$ in water, with respect to the protosolar value in hydrogen, by integrating the equation of diffusion within the solar nebula, as a function of the heliocentric distance and time. They compared the obtained value of $f$ to observations and determined a range of possible values for $M_{D0}$, $R_{D0}$, and $\alpha$. Better constraints on these parameters were obtained by Mousis~et~al.~(2000) who reunified the D/H ratio in proto-Uranian and proto-Neptunian ices with the cometary value by taking into account the recent interior models of Uranus and Neptune calculated by Podolak et al. (2000). Afterwards, Mousis (2004) employed the minimum mass and maximum mass solar nebula models defined by Mousis et al. (2000) in order to constrain the range of values of the D/H ratio in H$_2$O in the feeding zones of Jupiter and Saturn. Figure 3 is derived from his work and represents the evolution of the deuterium enrichment $f$ in H$_2$O in icy grains at the epoch of their formation in the solar nebula, as a function of the heliocentric distance for the two extreme mass solar nebula models. We assume that $f(R) = 31$ at $t = 0$ for D/H in water. This value corresponds to that measured in the most enriched component (D/H~=~$(73~\pm~12)~\times~10^{-5})$ found in LL3 meteorites (Deloule et al. 1998) which compares to a protosolar value assumed to be equal to $(2.35 \pm 0.3) \times 10^{-5}$ (Mousis et al. 2002b). It can be seen that icy grains formed at 5 AU in the solar nebula provide a deuterium enrichment $f$ which is between 3.8 and 4.7 times the protosolar value. With a (D/H)$_{proto}$ equal to $(2.35 \pm 0.3) \times 10^{-5}$, the corresponding D/H ratio in H$_2$O ice is between $7.8~\times~10^{-5}$ and $12.4~\times~10^{-5}$. In the case of the maximum mass solar nebula, icy particles formed between 5 and 9.8 AU also present D/H ratios in H$_2$O which are almost enclosed by these values. Hence, the subsequent $in~situ$ measurement of a D/H ratio in H$_2$O ice ranging between $7.8~\times~10^{-5}$ and $12.4~\times~10^{-5}$ on the surface of Ceres would imply it incorporated some icy planetesimals agglomerated from grains formed approximatively between 5 and 9.8 AU from the Sun. On the other hand, the $in~situ$ measurement of a D/H ratio higher than $12.4~\times~10^{-5}$ would mean that some icy planetesimals accreted by Ceres agglomerated from a mix of grains produced at a higher heliocentric distance in the solar nebula.

\begin{figure}
\centering
\includegraphics[width=6.3cm,clip=,angle=-90]{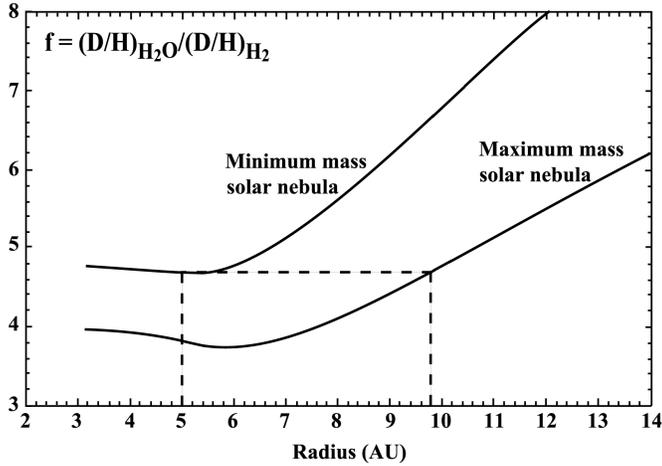}
 \caption{Deuterium enrichment factor $f$ as a function of the heliocentric distance in AU at the epochs of the condensation of water for the minimum mass and maximum mass solar nebula models derived from Drouart et al. (1999). The intersection of the left vertical dashed line with the two extreme mass solar nebula enrichment profiles gives the range of values of $f$ inferred in the icy grains produced at 5 AU. The intersection of the right vertical dashed line with the maximum mass solar nebula enrichment profile gives the value of $f$ at 9.8 AU. This value is similar to the one calculated at 5 AU with the minimum mass solar nebula (adapted from Mousis 2004).}
\label{FigVibStab}
\end{figure}

   \section{Summary and discussion}
We have shown that icy planetesimals produced beyond 5 AU can drift inward until the present location of Ceres and may be accreted by the forming asteroid. Moreover, we have demonstrated that the ices included in these planetesimals can be preserved during their radial drift. We have argued that volatiles were trapped in the form of clathrate hydrates, hydrates or pure condensates in the outer solar nebula prior to having been incorporated in icy planetesimals. This allowed us to calculate the per mass abundances with respect to H$_2$O of CO$_2$, CO, CH$_4$, N$_2$, NH$_3$, Ar, Xe, and Kr in the interior of Ceres, assuming that these volatiles were not vaporized during the accretion phase and the thermal history of the asteroid. The amount of ammonia calculated for Ceres is in agreement with the hypothesis invoked by McCord \& Sotin (2005) concerning the existence of a deep salty ocean in its current interior. Indeed, only a small amount of ammonia is required to lower the melting temperature of water in Ceres during its thermal evolution and may lead to the preservation of its deep liquid layer. The $Dawn$ space mission may have the capacity to test for the presence of a deep ocean in the interior of Ceres by making gravity measurements. Moreover, the presence of ices on the surface of Ceres, as well as their composition, is susceptible to be confirmed by the $Dawn$ mission. We have also shown that an $in~situ$ measurement of the D/H ratio in H$_2$O ice on the surface of Ceres would allow the zone of formation in the solar nebula of the volatile rich planetesimals that ultimately took part in its accretion to be assessed.\\

Our study is based on the assumption that Ceres contains a significant amount of volatiles in its interior. This has been first suggested by the simulations of Cyr et al (1998) and confirmed by our own calculations. This assumption may be questionable since it's noteworthy that no meteorite has been found to be completely aqueous altered. This suggests that water ice, as well as other volatiles, that were initially incorporated in parent bodies, were used up during aqueous alteration. On the other hand, as stated in Russel et al. (2004), one must note that no meteorite evidence directly linked to Ceres has been found. Three other clues are then in favor of the presence of ices in the interior of Ceres. First, according to the internal structure models of McCord \& Sotin (2003), the low density of the asteroid suggests that about 20 \% of its total mass is made of ices. However, one must mention that this low density may also be explained by alternative internal structure models. For instance, given the bulk densities of carbonaceous chondrites that could be contained in Ceres, the overall measured density of the asteroid may be reduced by associating it to a regolith layer with an important porosity. Second, a tenuous atmosphere has been detected around Ceres and this suggests the presence of ices on its surface (A'Hearn~\&~Feldman 1992). Finally, the recent Ceres observations with HST conducted by Parker~et~al.~(2004) tend to support the hypothesis of an internal structure containing a rocky core surrounded by a large icy shell. These evidences are then in favor of a significant content in volatiles in Ceres. However, the final answer to this important question will certainly be given by the $Dawn$ space mission, that  will also bring clues on the different physical processes that occured during the formation of the main belt.

\section*{Acknowledgments}

This work was supported in part by the Swiss National Science Foundation. OM was supported by an ESA external fellowship, and this support is gratefully acknowledged. Many thanks are due to Keith Holsapple and Jonathan Horner for their valuable comments on the manuscript. We acknowledge Christophe Sotin for enlightning discussions and information on his work.

\label{lastpage}

\end{document}